\documentclass [12pt]{article}
\usepackage{graphicx}
\usepackage{amsmath}
\usepackage{indentfirst}

\begin{document}

\begin{center}
\textbf{\Large The dark energy effect as manifestation of the signal
 propagation features in expanding Universe}
\end{center}

\begin{center}
L.M.Tomilchik
\end{center}

\begin{center}
B.I. Stepanov Institute of Physics, National Academy \\ of
Sciences of Belarus, Minsk, Belarus.
\end{center}

\begin{abstract}
The formula describing the explicit red-shift dependence for the
distance covered by a signal in the expanding space, and also the
expressions for Hubble law have been derived directly from the
conformal group transformations taking into account the light-cone
equation.  The obtained relations are associated with the
luminosity distance represented as an explicit function of the red
shift involving no other parameters but the scaling factor $c
H^{-1}_0$. These relations represent in a pure kinematical way and
in a complete agreement with observations, the revealed transition
of the Metagalaxy to accelerated expansion (the dark energy
effect).
\end{abstract}

\section{Introduction.}

Modern observational astronomy, apart from yielding a series of sensational
experimental findings, has posed serious problems for the conventional
cosmology, and fundamental interaction physics as a whole. So, an
interpretation of the revealed effect of the accelerated Universe expansion
[1,2] within the scope of a conventional cosmological model has required the
use of such equations of state for the gravitating substance (dark energy,
quintessence) which are not in line with any of the observable types.
However, just this substance should make the major part (up to three
fourths) of the total energy content in the Universe. Such a situation is
liable to stimulate the development of alternative approaches including the
efforts to modify the dynamic principles of a gravitation theory (see, for
example, [3,4] and references herein).

But the well-known historic precedent with the space filling by unobservable
luminiferous ether suggests that, trying to find dynamic explanations for
the ``strange'' observable effects of new kinematics, the researchers
inevitably come across serious difficulties and paradoxes.

Let us recollect the effect of the accelerated Universe expansion,
purely from the viewpoint of observation. In the fundamental
experiments realized using the $Ia$ - type supernova as standard
candles it has been found that over the interval of the
cosmological red shift $0.2<z<1.6$~there are specific deviations
from a linear Hubble law. They are characterized by a marked
maximum close to $z_{\max } \approx 0.5$ $ \left( {z_{\max
}^{\exp } = 0.46\pm 0.13} \right)$ [1,2] and show tendency to
monotonic decrease [5]. Aside from pure photometric parameters,
the principal quantity directly recorded during the experiments is
a frequency shift in the electromagnetic signals arriving to the
observation point from recessing sources. And it is only natural
to assume that the observed phenomenon is mainly caused by the
propagation features of free electromagnetic radiation in the
expanding space, i.e. the effect may be kinematic rather than
dynamic in its nature. Because of this, we should look for the key
to its explanation in the kinematic principles of cosmology.

Unfortunately, the problem of isolating the kinematic component in
the structure of GR is far from unambiguity characteristic for
Spetial Relativity (SR). In fact, it is a matter of some or other
geometry selection for the background space-time manifold, i.e.
space-time in the absence of gravitating sources and of a
gravitation field per se (``empty'' Universe). Within the scope of
a conventional approach, such a geometry is identified (locally)
with that of the Minkowski space, where the corresponding group of
isometries offers invariance of the light-cone equation as one of
the basic kinematic principles of SR (synchronization condition of
the spatially separated mutually stationary clocks).

In conditions of the spatial curvature approximating zero, it seems
reasonable to retain the key invariance condition for the light cone
equation in the case of the nonstationary spatial scales as well. It is
commonly known that an extension of the Minkowski-space symmetry group
meeting this condition is the conformal group SO(4,2) including, aside from
isometries, the transformations which modify the space-time scales.

In addition to Lorentz transformations (LT) as a subgroup, the group SO(4,2)
also includes a group of special conformal transformations (SCT). The metric
form of Minkowski space

\[
S^2 = x^\mu x_\mu = \left( {x^2} \right),\quad \left( {g_{\mu \nu }
= diag\left\{ {1, - 1, - 1, - 1} \right\}} \right),
\]

\noindent
is invariant relative to LT

\begin{equation}
\label{eq1}
{x}'^\mu = L_\alpha ^\mu x^\alpha \quad \left( {L_\alpha ^\mu L_\nu ^\alpha
= \delta _\nu ^\mu } \right),
\end{equation}

\noindent
i.e. the following equality is valid:

\[
\left( {{x}'^2} \right) = \left( {x^2} \right).
\]

As regards SCT of the form

\begin{equation}
\label{eq2}
{x}'^\mu = \sigma ^{ - 1}\left( {a,x} \right)\left\{ {x^\mu + a^\mu \left(
{x^2} \right)} \right\},
\end{equation}

\noindent
where

\begin{equation}
\label{eq3}
\sigma \left( {a,x} \right) = 1 + 2\left( {ax} \right) + \left( {a^2}
\right)\left( {x^2} \right),
\end{equation}

\noindent $a^\mu $ -- four-vector parameter, the metric Lorentz
invariant is multiplied by the Lorentz-invariant
coordinate-dependent factor, and we have

\begin{equation}
\label{eq4}
\left( {{x}'^2} \right) = \sigma ^{ - 1}\left( {a,x} \right)\left( {x^2}
\right).
\end{equation}

But as seen from (\ref{eq3}), with regard to $\sigma ^{ - 1}\left.
{\left( {a,x} \right)} \right|_{x^2 = 0} \ne 0,$ in both cases the
conditions $\left( {{x}'^2} \right) = \left( {x^2} \right) = 0$ are
mutually determined. This is a well known case of the light-cone
equation invariance both with respect to LT and SCT.

From the viewpoint of conformal geometry, the propagation of
electromagnetic signals in the case of expansion must follow the
same laws as in Minkowski space-time, i.e. should be realized at a
uniform speed along the light-cone generators in each pair of the
reference systems (RS) selected in the background space-time and
related by LT and/or by SCT.

An experimental checking of the cosmological expansion law
presupposes an independent determination of the galactic objects
recession speed and of the distance from each of the objects to the
observation point. In so doing the most informative and exactly
measurable quantity is a frequency shift in the arriving
electromagnetic signals. A theoretical description of the expansion
fitting the observations adequately may be obtained with the
expressions defining the explicit frequency-shift dependence for a
relative velocity of the radiation source and detector as well as
for the distance covered by the signal from the moment of its
emission to the observation time. The formula relating the ray
relative velocity V$_{R}$ of the source and detector to the red
shift $z$ is well known from a relativistic theory of Doppler effect

\begin{equation}
\label{eq5}
V_R \left( z \right) = c\frac{\left( {1 + z} \right)^2 - 1}{\left( {1 + z}
\right)^2 + 1}.
\end{equation}

\noindent This formula remains applicable with the expansion too.

In this work the formula describing the explicit red-shift
dependence for the distance covered by a signal in the expanding
space, and also the expressions for Hubble law have been derived
directly from the conformal group transformations taking into
account the light-cone equation, i.e. in a pure kinematic way. It
has been found that the obtained relations are associated with the
luminosity distance represented as an explicit function of the red
shift involving no other parameters but the scaling factor $c
H^{-1}_0$. These relations represent, in a complete agreement with
observations, the revealed transition of the Metagalaxy to
expansion, making it possible to verify experimentally the
predictions concerning the observable deviations of Hubble law
from a linear behavior over a wide interval of the red shift.

\section{ Distance as a function of the cosmological red shift and Hubble
law }

First, let us present an elementary derivation of the formulae, expressing
the red-shift dependence of the recession speed and of the distance covered
by a signal in conditions of the expanding Universe, directly from the
conformal group transformations.

Since within a spatially homogeneous and isotropic cosmological model a
leading role is played by the longitudinal motion component, for simplicity,
we consider the case of one spatial dimension to define the four-vector
$x^\mu $ as follows:

\begin{equation}
\label{eq6} x^\mu = \left\{ {x_0 = ct,x,0,0} \right\}.
\end{equation}

The vector of a relative velocity (parameter of a group of Lorentz boosts)
having a single component is given by

\begin{equation}
\label{eq7}
{V} = \left\{ {V_R ,0,0} \right\}.
\end{equation}

The four-vector $a^\mu $ of the group in SCT (formulae (\ref{eq2}), (\ref{eq3})), in
accordance with [6], will be taken in the form

\begin{equation}
\label{eq8}
a^\mu = \left\{ {0, - \frac{1}{2ct_{\lim } },0,0} \right\},
\end{equation}

\noindent where the parameter $t_{\lim } $ has the dimension of
time. In our consideration it is convenient to use the light-cone
variables

\begin{equation}
\label{eq9}
u = \frac{x_0 + x}{2},\quad v = \frac{x_0 - x}{2}.
\end{equation}

Then LT(\ref{eq1}) and SCT(\ref{eq2}), (\ref{eq3}) may be written as

\begin{equation}
\label{eq10}
{u}' = \left( {\frac{1 - \beta }{1 + \beta }} \right)^{\frac{1}{2}}u,\quad
{v}' = \left( {\frac{1 + \beta }{1 - \beta }} \right)^{\frac{1}{2}}v,
\end{equation}

\noindent
where $\beta = \raise0.7ex\hbox{${V_R }$} \!\mathord{\left/ {\vphantom {{V_R
} c}}\right.\kern-\nulldelimiterspace}\!\lower0.7ex\hbox{$c$}$;

\begin{equation}
\label{eq11} {u}' = \frac{u\left( {1 - \cfrac{v}{ct_{\lim } }}
\right)}{1 + \cfrac{u - v}{ct_{\lim } } - \cfrac{uv}{c^2t_{\lim }^2
}},\quad {v}' = \frac{v\left( {1 + \cfrac{u}{ct_{\lim } }}
\right)}{1 + \cfrac{u - v}{ct_{\lim } } - \cfrac{uv}{c^2t_{\lim }^2
}}.
\end{equation}

An expression for the interval $S^2 = x^\mu x_\mu $ is of the form

\begin{equation}
\label{eq12}
S^2 = 4uv.
\end{equation}

The light-cone equation $S^2 = 0$ is written as

\begin{equation}
\label{eq13}
uv = 0.
\end{equation}

This equation has the following solutions:

\begin{equation}
\label{eq14}
v = 0,\quad u \ne 0;
\end{equation}

\begin{equation}
\label{eq15}
u = 0,\quad v \ne 0.
\end{equation}

For the variables $x$, $t$ these solutions indicate the
fulfillment of the relations in the cases (\ref{eq14}) and
(\ref{eq15}) correspondingly

\[
x = ct,\quad u = x = ct;
\]

\[
x = - ct,\quad v = - x = ct;
\]

\noindent i.e. (\ref{eq14}) ((\ref{eq15})) conforms to the signal
propagation in the positive (negative) direction of the light-cone
generators.

Next we use LT defined by formulae (\ref{eq10}) with due regard for (\ref{eq14}) and (\ref{eq15}) to
find (the upper and lower signs in formulae hereinafter are associated with
the signal propagation in the forward and backward directions) that

\begin{equation}
\label{eq16}
{t}' = \left( {\frac{1 \mp \beta }{1\pm \beta }} \right)^{\frac{1}{2}}t.
\end{equation}

From the viewpoint of geometry, transformation (\ref{eq16})
represents a linear deformation of the light-cone generators for
LT. In a similar way, with regard to (\ref{eq14}), (\ref{eq15}),
from formulae (\ref{eq2}) and (\ref{eq3}) we obtain

\begin{equation}
\label{eq17}
{t}' = \left( {1\pm \frac{t}{t_{\lim } }} \right)^{ - 1}t.
\end{equation}

Geometrically, (\ref{eq17}) denotes a nonlinear deformation of the
light-cone generators -- conformal transformation of a
semi-infinite line to the finite length t$_{lim}$ (upper sign) and
the inverse transformation (lower sign).

From transformations (\ref{eq16}) and (\ref{eq17}) we get the following formulae for
differentially small increments in time as follows:

\begin{equation}
\label{eq18}
\Delta {t}' = \left( {\frac{1 \mp \beta }{1\pm \beta }}
\right)^{\frac{1}{2}}\Delta t,
\end{equation}

\begin{equation}
\label{eq19}
\Delta {t}' = \left( {1\pm \frac{t}{t_{\lim } }} \right)^{ - 2}\Delta t.
\end{equation}

Taking $\Delta t$ and $\Delta {t}'$ as oscillation periods of the
emitted (~T$_{emitted}$~) and observed (arriving to the observation
point) (~T$_{observed}$~) signals and considering the relationship
between the wavelength and oscillation period, we can derive the
following expressions for the red shift case (lower signs in
(\ref{eq18}) and (\ref{eq19})) of interest:

\begin{equation}
\label{eq20}
\raise0.7ex\hbox{${\lambda _{obs} }$} \!\mathord{\left/ {\vphantom {{\lambda
_{obs} } {\lambda _{em}
}}}\right.\kern-\nulldelimiterspace}\!\lower0.7ex\hbox{${\lambda _{em} }$} =
\left( {\frac{1 + \beta }{1 - \beta }} \right)^{\frac{1}{2}} = z + 1,
\end{equation}

\begin{equation}
\label{eq21}
\raise0.7ex\hbox{${\lambda _{obs} }$} \!\mathord{\left/ {\vphantom {{\lambda
_{obs} } {\lambda _{em}
}}}\right.\kern-\nulldelimiterspace}\!\lower0.7ex\hbox{${\lambda _{em} }$} =
\left( {1 - \frac{t}{t_{\lim } }} \right)^{ - 2} = z + 1.
\end{equation}

Solving relation (\ref{eq20}) with respect to $\beta $, we obtain
expression (\ref{eq5}) for the velocity $V_R \left( z \right)$. From
(\ref{eq21}) we can derive the following formula for the signal
propagation time t(z):

\begin{equation}
\label{eq22}
t\left( z \right) = t_{\lim } \left\{ {1 - \left( {1 + z} \right)^{ -
\frac{1}{2}}} \right\}.
\end{equation}

Taking into consideration a vanishingly small curvature of the
modern Metagalaxy, we can find the  distance covered by the signal
as

\begin{equation}
\label{eq23}
D\left( z \right) = ct\left( z \right) = ct_{\lim } \left\{ {1 - \left( {1 +
z} \right)^{ - \frac{1}{2}}} \right\}.
\end{equation}

The parameter $t_{lim}$ is easily related to Hubble constant with
regard to the fact that for small values of the red shift the
experimentally observed dependence of the galactic recession speed
on the intergalactic distance is linear $\left( {V = H_0 D}
\right)$ to a high accuracy. As seen from formulae (\ref{eq5}) and
(\ref{eq23}), the expressions for $V_R \left( z \right)$ and
$D(z)$ at $z \ll 1$ are proportional to each other up to the
factors $\sim z^2$. Indeed, from (\ref{eq5}) and (\ref{eq23}) we
derive

\[
\left. {\raise0.7ex\hbox{${V_R \left( z \right)}$} \!\mathord{\left/
{\vphantom {{V_R \left( z \right)} {D\left( z
\right)}}}\right.\kern-\nulldelimiterspace}\!\lower0.7ex\hbox{${D\left( z
\right)}$}} \right|_{z \ll 1} \cong 2t_{\lim }^{ - 1} ,
\]

\noindent
whence it follows that $t_{\lim } = 2H_0^{ - 1} .$

The expression for Hubble law written as

\begin{equation}
\label{eq24}
\raise0.7ex\hbox{${V_R \left( z \right)}$} \!\mathord{\left/ {\vphantom
{{V_R \left( z \right)} {H_0 D\left( z
\right)}}}\right.\kern-\nulldelimiterspace}\!\lower0.7ex\hbox{${H_0 D\left(
z \right)}$} = f\left( z \right) = \frac{1}{2}\frac{\left( {1 + z}
\right)^{\frac{1}{2}}}{\left( {1 + z} \right)^2 + 1} \cdot \frac{\left( {1 +
z} \right)^2 - 1}{\left( {1 + z} \right)^{\frac{1}{2}} - 1}.
\end{equation}

\noindent is determined over the whole variation interval of $z$.

In the approximation of $z\ll1$, it represents a linear form of
Hubble law $\left( {V_R = H_0 D} \right)$. The function $f(z)$
possess a maximum at $z_{\max } \cong 0.474$, that surprisingly
well agrees with a position of the experimentally found maximum
for $z_{\max }^{\exp } = 0.46\pm 0.13$ interpreted as a transition
point from decelerated to the accelerated expansion of the
Universe [1, 2].

Yet, a direct comparison of expressions (\ref{eq22}) and
(\ref{eq24}) with the experimental data presents difficulties. It
is a standard practice in astronomy to use for description of the
observable dependences the so-called luminosity distance given by

\begin{equation}
\label{eq25}
d_L = \left( {\frac{L}{4\pi F}} \right)^{\frac{1}{2}},
\end{equation}

\noindent where $L$ -- luminosity of a distant radiation source,
$F$ -- measurable energy flux from the source at the observation
point.

On the other hand, in modern cosmology the basic relation relating the
galactic red shift to the space-time characteristics of the expanding
Universe is as follows (e.g., see [7]):

\begin{equation}
\label{eq26}
\frac{a\left( {t_{fin} } \right)}{a\left( {t_{in} } \right)} = z + 1.
\end{equation}

Here $a\left( {t_{in} } \right)$ and $a\left( {t_{fin} } \right)$
are the values of the scaling factor a(t) in the preceding
($t_{in})$ and subsequent ($t_{fin})$ instants of time, $z$ -- red
shift. Using the distance determined as $D=ct$, we can write
expression (\ref{eq26}) in the following form:

\begin{equation}
\label{eq27}
\frac{a\left( {t_{fin} } \right)}{a\left( {t_{fin} - \frac{D}{c}} \right)} =
z + 1.
\end{equation}

To associate expressions (\ref{eq26}), (\ref{eq27}) with the
observables, one can use Taylor expansion of the scaling factor in
terms of time (actually, in terms of H$_{0}$t) or $D/c$, taking all
time derivatives at the time of observation $t_0 = t_{fin}$. Note
that only two first terms of the expansion involve the quantities
well grounded within the scope of a conventional cosmological
Friedmann--Robertson--Walker (FRW) model  and included into the
dynamic equations of Friedmann. These quantities are (i) Hubble
parameter determining the expansion rate and given as

\begin{equation}
\label{eq28}
H\left( t \right) = a^{ - 1}\left( t \right)\mathop a\limits^ \cdot \left( t
\right),
\end{equation}

\noindent (ii) deceleration parameter determining changes in this
rate and given by the formula

\begin{equation}
\label{eq29}
q\left( t \right) = - a^{ - 1}\left( t \right)H^{ - 2}\left( t
\right)\mathop a\limits^{ \cdot \cdot } \left( t \right).
\end{equation}

In FRW - model this parameter relates the sign and value of
acceleration to the pressure and energy density of the gravitating
matter. A dot over $a(t)$ denotes differentiation in time. As seen
from definition of (\ref{eq29}), a positive (negative) value of
q(t) is associated with decelerated (accelerated) expansion. The
values of $H(t)$ and $q(t)$ taken in the time of observation
$t_{fin} = t_0 $, represent the measured Hubble constant H$_{0}$
and deceleration parameter $q_{0}$, respectively.

Restriction to the first two terms in Taylor expansion turned to
be sufficient for the fulfillment of the condition $z\ll1$. When
proceeding to the region of greater values of $z$, one should take
into consideration further expansion terms leading to new,
phenomenological, parameters (jerk, snap, etc.) associated with
time derivatives of the scaling factor higher than the
second-order (see [5, 8] and references herein).

Within the scope of a conventional Friedmann cosmology, all
expressions characterizing the space-time properties of the
expanding Universe are also dependent on $H_{0}$ and $q_{0}$ in
addition to the variable $z$. For example, an approximate formula
for the signal propagation time, including the terms quadratic in
$z$, is of the form [7]

\begin{equation}
\label{eq30}
t\left( {H_0 ,q_0 ,z} \right) = H_0^{ - 1} \left\{ {z - \left( {1 +
\frac{q_0 }{2}} \right)z^2 + O\left( {z^3} \right)} \right\}.
\end{equation}

In FRW - model the deceleration parameter is related to a
dimensionless density of the matter $\Omega =
\raise0.7ex\hbox{$\rho $} \!\mathord{\left/ {\vphantom {\rho {\rho
_c }}}\right.\kern-\nulldelimiterspace}\!\lower0.7ex\hbox{${\rho
_c }$},$ where $\rho $ -- mean energy density in the Universe and
$\rho _{c}$ -- critical density, by means of the well-known
relation

\begin{equation}
\label{eq31}
q_0 = \frac{\Omega }{2}\left( {1 + 3w} \right).
\end{equation}

\noindent Here $w$ -- numerical factor relating the pressure $p$
and energy density $\rho $ in the equation of state

\begin{equation*}
p = w\rho .
\end{equation*}

\noindent Assuming the observable acceleration to be a result
(outcome) of total influence from (of) separate types of the
gravitating substance and neglecting the contribution from (of)
the relativistic gas $\left(w=\frac{1 }{3}\right)$ let us write
according (\ref{eq31})

\begin{equation}
\label{eq32} q_0 = \frac{\Omega_M }{2}\left( {1 + 3w_M}
\right)+\frac{\Omega_\Lambda }{2}\left( {1 + 3w_\Lambda}
\right)=\frac{\Omega_M }{2}-\Omega_\Lambda,
\end{equation}

\noindent where $\Omega_M$ is cold matter density, including the
dark one ($w=w_M=0$), $\Omega_\Lambda$ is dark energy density
($w=w_\Lambda=-1$).

To compare formula (\ref{eq22}) to expression (\ref{eq30}), let us
perform a series expansion of the function $t(z)$ limiting
ourselves to the term quadratic in $z$. As a result, we have

\begin{equation}
\label{eq33}
t\left( z \right) = H_0^{ - 1} \left\{ {z - \frac{3}{4}z^2 + O\left( {z^3}
\right)} \right\}.
\end{equation}

It is obvious that expressions (\ref{eq30}) and (\ref{eq33}) are
coincident for $q_0 = - \frac{1}{2}$. Because of a vanishingly
small spatial curvature of the modern Metagalaxy the mean matter
density is practically coinciding with the critical one. Therefore
we can assume

\begin{equation*}
\Omega=\Omega_M +\Omega_\Lambda =1.
\end{equation*}

\noindent  Adding the equation (\ref{eq32}) under condition $q_0 =
- \frac{1}{2}$, i.e.

\begin{equation*}
2\Omega_\Lambda-\Omega_M  =1
\end{equation*}

\noindent we can see that

\begin{equation*}
\Omega_M  =\frac{1}{3},\quad \Omega_\Lambda=\frac{2}{3}.
\end{equation*}

\noindent These numerical magnitudes are near to ones which found
experimentally ($\Omega_M\approx0.3,
\quad\Omega_\Lambda\approx0.7$) [5]

 It is seen that the exact formula given by (\ref{eq22}) for
$t(z)$, derived proceeding from pure kinematic considerations, in
the approximation under study represents the result that in a
dynamic approach of GR is associated with a negative value of the
deceleration parameter (i.e. accelerated expansion) and hence with
a negative pressure in state equation  (dark energy \linebreak
effect [1, 2]).

The dependence given by formula (\ref{eq22}) arise in the case of
FRM-type metric with the scale factor $a(t)=\left( {\frac{t}{t_0}}
\right)^{2/3}\quad \left(t_0= 2H_0^{ - 1}/3,\quad a(t_0)=1 \right)$
which corresponds to the spacially-flat Universe with $w=0$ (see for
example [13]). But this metric leads to purely decelerating
expansion $(\Omega =\Omega_M =1,\quad q_0=1/2 )$. It can be easily
seen that just the same dependence $t(z)$ can be obtained in the
case $a(t)=\left( 1+\frac{t}{t_0} \right)^{2} \quad \left(t_0=
2H_0^{ - 1},\quad a(0)=1 \right)$ too. Such expression gives correct
picture of the nowadays observable expansion (the deceleration
parameter is equal to $ - 1/2$).

\section{ Luminosity distance as a function of cosmological red shift.}

A standard practice in modern astrophysics is to represent the experimental
Hubble diagrams, giving the dependence of the luminosity distance d$_{L}$ on
the red shift, as a function $\mu \left( {d_L } \right)$ (distance modulus)
of the following form (see [5, 7]):

\begin{equation}
\label{eq34} \mu = 5\log \left( {\frac{d_L }{Mpc}} \right) + 25,
\end{equation}

\noindent
where d$_{L}$ is measured in megaparsecs (Mpc).

It should be noted that all the available expressions for the
luminosity distance derived within a conventional approach of GR
are dependent not only on the variable z but also on the
additional varied parameters (see [5] and references herein).

Now our problem is to express the luminosity distance d$_{L}$ in
terms of the distance D that is covered by a light signal, emitted
in the remote past, to the moment of its recording at the
observation point. In the absence of the closed-form analytical
expressions relating the distances d$_{L}$ and D,  we can use the
representation of d$_{L}$ derived  in the form of Taylor expansion
in powers of D up to the fourth-order terms [8].

We limit ourselves to the first two expansion terms since further terms
(beginning from the third) include, apart from the Hubble constant H$_{0}$
and deceleration parameter q$_{0}$, the additional phenomenological
parameters (jerk, snap).

The corresponding expression is of the form (see [8], formula (45))

\begin{equation}
\label{eq35}
d_L \left( D \right) = D\left\{ {1 + \frac{3}{2}\left( {\frac{H_0 D}{c}}
\right) + \frac{1}{6}\left[ {11 + 4q_0 - \frac{kc^2}{H_0^2 a_0 }}
\right]\left( {\frac{H_0 D}{c}} \right)^2} \right\}.
\end{equation}

\noindent Here $k = 0,\pm 1$ -- curvature index, a$_{0}$=a(t$_{0})$
-- current value of the scaling factor. As the relation d$_{L}$(D)
is general in character, by substitution of $D = D\left( z \right)$,
where the distance D(z) is determined by (\ref{eq23}), we can obtain
an expression for the luminosity distance as a function of the
variable z, that involves no other parameters except the common
scaling factor $cH_0^{ - 1} $. In the process we make allowance for
the fact that the Metagalaxy is spatially Euclidean (k=0), also
assuming $q_0 = - \frac{1}{2}$ in accordance with the foregoing.

As a result, we have the following explicit function $d_L \left( z
\right)$:

\begin{equation}
\label{eq36}
d_L \left( z \right) = 2cH_0^{ - 1} \left\{ {\Phi \left( z \right) + 3\Phi
^2\left( z \right) + 6\Phi ^3\left( z \right)} \right\},
\end{equation}

\noindent
where

\begin{equation}
\label{eq37}
\Phi \left( z \right) = 1 - \left( {1 + z} \right)^{ - \frac{1}{2}}.
\end{equation}

\noindent Specifying for H$_{0}$ the numerical value $H_0 = 73 \;
km\left( {Mpc} \right)^{ - 1}s^{ - 1} \cong 2.37 \cdot 10^{ - 18}\;s^{
- 1}$, we can find $2cH_0^{ - 1} \cong 0.82 \cdot 10^4\; Mpc$.

Then expression (\ref{eq34}) for the distance modulus $\mu \left(
z \right)$ may be rewritten as

\begin{equation}
\label{eq38} \mu \left( z \right) = 5\log \left\{ {0.82 \cdot
10^4\left[ {\Phi \left( z \right) + 3\Phi ^2\left( z \right) +
6\Phi ^3\left( z \right)} \right]} \right\} + 25,
\end{equation}

\noindent
where $\Phi \left( z \right)$ is determined by (\ref{eq37}).

The function $\mu $(z) is graphically shown in Fig.~1.

\begin{figure}[h]
\centering
\includegraphics[width=2.00in]{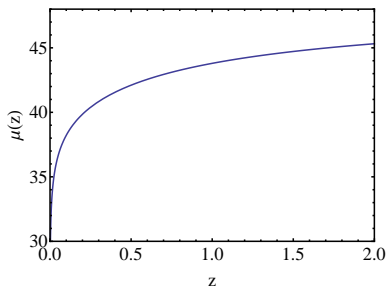}
\caption{\small The function $\mu $(z) over the interval from 0.0
to 2.0.}
\end{figure}

\begin{figure}[h]
\centering
\includegraphics[width=2.00in]{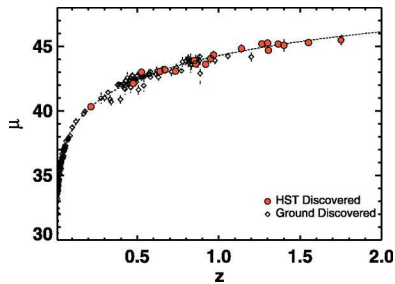}
\caption{\small Experimental Hubble diagram.}
\end{figure}

For comparison, Fig.~2 shows the experimental Hubble diagram
borrowed from the original work [5]. As seen, the curve calculated
on the basis of (\ref{eq38}) represents the experimentally observed
function $\mu _{exp} \left( z \right)$ over the interval $0<z<2$
with a good accuracy.

For a more detailed comparison with the experimental data, let us consider
the expression

\begin{equation}
\label{eq39}
\delta _f \left( z \right) = 5\left\{ {f\left( z \right) - 1} \right\},
\end{equation}

\noindent where the function f(z) is determined by (\ref{eq24}).
This expression defines deviation of Hubble law from a strictly
linear behavior over the whole range of the red shift values. The
multiplier 5 is used because this comparison is performed with the
experimental data expressed in terms of $\Delta \mu $, where $\mu $
is given by (\ref{eq34}).

Note that expressions (\ref{eq22}) and (\ref{eq24}) used for definition of the function in
(\ref{eq39}) have been obtained in the assumption of time independence for the
Hubble parameter. Because of this, the corresponding formulae are obviously
inapplicable for very great values of z. But we can assume that, for
certain, their applicability region covers the red shifts on the order of
several units.

\begin{figure}[h]
\centering
\includegraphics[width=2.00in]{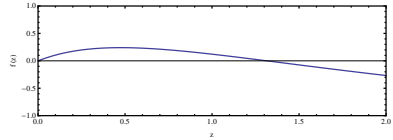}
\caption{\small Deviations from linear Hubble law.}
\end{figure}

\begin{figure}[h]
\centering
\includegraphics[width=2.00in,height=1.25in]{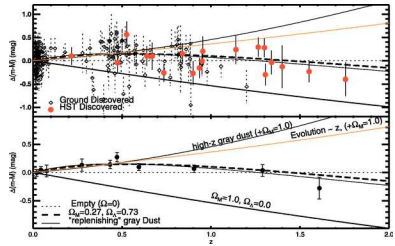}
\caption{\small Residual Hubble diagram.}
\end{figure}

The function $\delta \left( z \right)$ over the interval $0<z<2$ is
graphically shown in Fig.~3. For comparison, Fig.~4 gives the
experimental residual Hubble diagram borrowed from [5] (see Fig.~7
there) and representing the observed dependence of deviations from a
linear Hubble law in the corresponding range of the measured
red-shift values.

Here the diamonds and circles denote experimental points, the
circles being associated with the red shifts in signals from sixteen
supernovae (SNe Ia) detected during a period from 2002 to 2003 (see
[5]). The full circles at the bottom of this figure denote the mean
weighted values calculated for several selected intervals of red
shifts. A zero dotted line in both parts is associated with an
``empty'' model of the Universe $\left( {\Omega = 0} \right)$and
with the approximation linear in $z$. The dashed line giving the best
fit to the experimental data has been obtained for $\Omega _M =
0,27,\;\Omega _\Lambda = 0,73,$ i.e. in the assumption that the
total $\left( {\Omega _M } \right)$ of the ordinary and cold dark
matter comes to 27{\%}, whereas a share of the dark energy $\left(
{\Omega _\Lambda } \right)$ amounts to 73{\%} of the overall
quantity of the gravitating substance in the modern Universe. The
solid thin lines above and below the zero line have been obtained
within the scope of some alternative approaches, the predictions of
which are at variance with the experimental data.

As seen, the curve (\ref{eq39}) shown in Fig.~3 represents well not
only the maximum position and value close to the point $z \approx
0.5$ but also the general form (behavior) of the observed dependence
on $z$, including a marked tendency of the experimental points to
drift to the region of negative deviations for $z$ in excess of $z_0
\approx 1.3$ ($z_0 = 1.315$ -- intersection point of the curve
(\ref{eq39}) with the $z$-axis). As the amplitude and character of
deviations for z values beyond the limits of the currently available
experimental data are described by the explicit analytical function
(\ref{eq39}), the predicted further behavior of the associated curve
may be experimentally checked directly in the process of detecting
new, very remote, standard distance indicators (Ia--type
supernovae). It should be particularly emphasized that the formulae
proposed involve no free parameters and the predicted results are
model-independent. On the other hand, the experimentally observed
pattern is represented within a standard cosmological model of GR by
a choice of the appropriate combination of variable parameters whose
number is growing with the red shifts.

As demonstrated in [9], the observed deviations of Hubble law from a
strictly linear relationship between the recession speed and
cosmological distance are directly caused by different dependences
of the functions $V_R \left( z \right)$ and $D(z)$ on the red shift.
Over the experimentally studied interval ($0<z<2$), for $z$ values
this difference is minor (below 5{\%}). And it is interesting to
compare the predictions calculated by formula (\ref{eq39}) with
those obtained independently using the functions $D(z)$ and $V_R
\left( z \right)$ as the distance-determining expressions. The
corresponding expressions for deviations are written as

\begin{equation}
\label{eq40}
\delta _D \left( z \right) = 2\left\{ {1 - \left( {1 + z} \right)^{ -
\frac{1}{2}}} \right\} - \alpha z,
\end{equation}

\begin{equation}
\label{eq41}
\delta _V \left( z \right) = \frac{\left( {1 + z} \right)^2 - 1}{\left( {1 +
z} \right)^2 + 1} - \alpha z,
\end{equation}

\noindent where $\alpha = 0.521.$

Fig.~5 shows to a single scale the curves for $\delta _f \left( z
\right)$ (solid line) and also for $\delta_D \left( z \right)$
(dashed line) and $\delta _V \left( z \right)$ (dotted line)
multiplied by 5 over the interval $0<z<2$.

\begin{figure}[h]
\centering
\includegraphics[width=4in]{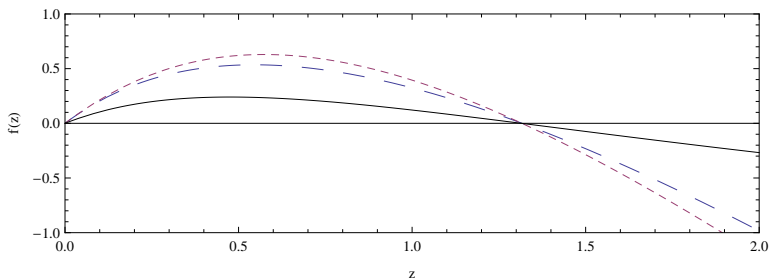}
\label{fig5}\caption{\small This figure shows to a single scale
the curves for $\delta _f \left( z \right)$ (solid line) and also
for $\delta _D \left( z \right)$ (dashed line) and $\delta _V
\left( z \right)$ (dotted line) multiplied by 5 over the interval
$0<z<2$.}
\end{figure}

All these three curves are similar in their structure: zero line
intersections (approximately) at the same point $z_0 \cong 1.315$;
maxima within the  \linebreak interval [0 -- 1.315]. However, from Table 1
it is seen that there are distinctions in the positions and values
of these maxima.

\begin{table}[h]
\centering
 \caption{The  positions of maxima and corresponding
values of the functions $\delta _f \left( z \right)$, $\delta _D
\left( z \right)$, $\delta _V \left( z \right)$.}\label{tab3}
{\begin{tabular}{|p{106pt}|p{92pt}|p{85pt}|}
\hline curve & z$_{max}$ & $\Delta $(z$_{max})$ \\
\hline $\delta _f \left( z \right)$ & 0.475 & 0.240 \\
\hline $\delta _D \left( z \right)$ & 0.544 & 0.535 \\
\hline $\delta _V \left( z \right)$ & 0.573 & 0.629 \\
\hline
\end{tabular}}
\end{table}

We can see that a qualitative and quantitative agreement with the
experiment when using formula (\ref{eq39}) is clearly worse when the
functions given by (\ref{eq40}) and (\ref{eq41}) are used
independently. This may be interpreted as a direct experimental
support for the expression (\ref{eq24}) derived in [9, 10] for the
cosmological expansion law.

\section{ Conclusion}

It has been demonstrated that the observable features of a modern
stage in the Universe expansion may be described consistently as
pure kinematic manifestations of the conformal space-time geometry.
A non-Euclidean character of the space-time background manifold in
this approach is represented by a conformal inhomogeneity of time.
Of course, such a theoretical model looks like the phenomenological
one. Nevertheless, the model not only represents well the observable
effects but also makes possible direct experimental checking with
further progress of the experiments into the region of higher
red-shift values. Even the fact that within this model the derived
explicit expressions for the cosmological expansion law and the
relationships between the distance and red shift require no variable
parameters for agreement with the observations suggests that this
model actually represents the real properties of the Metagalaxy at a
modern stage of its expansion.

Besides, as shown in [9] (see also [10,11] and the references
herein), with the use of expression (\ref{eq17}) determining a
conformal deformation of the light-cone generators we can,
qualitatively and quantitatively, represent in the approximation
$H_0 t \ll 1$ the observed Pioneer anomaly effect (anomalous
violet frequency shift in signals of the receding source) as a
universal local manifestation of the cosmological expansion, in
principle, detected at other frequencies (e.g., optical) as well.

It is important that the experimentally recorded local expansion effects may
be interpreted as an exhibition of nonintertiality of the observer's
reference system (RS), ``expanding RS'', in every point of which there is a
background acceleration directed to the observation point, its numerical
value being determined by the expansion rate (i.e. by the observable
Hubble-parameter value).

It is clear that, due to a pure kinematic character of the approach, in the
proposed expansion pattern the gravitating sources and fields themselves are
lacking. But gravity may be introduced on the basis of the equivalence
principle as an effective background field whose ``strength'', similar to
the corresponding background acceleration that is equal to cH$_{0}$, is
determined by Hubble constant.

The Hubble constant in the proposed model is used as a limiting
value of the parameter setting the fixed time scale that, together
with the speed of light, gives the space-time dimensions of the
observable Universe. In so doing the status of the constants c and
H$_{0}$ is differing considerably. The upper limit of speed is
fundamental in its character, being naturally involved in the
principles of SR. However, no arguments for the possibility of
bounds on a numerical value of Hubble constant as an inference
from some general physical principle have been found. This problem
necessitates special attention. It seems that most promising in
this respect looks the Maximum Tension Principle put forward by
Gibbons [12]. As demonstrated by the estimates including a maximal
force after Gibbons ($F_{\max } = \raise0.7ex\hbox{${c^4}$}
\!\mathord{\left/ {\vphantom {{c^4}
{4G}}}\right.\kern-\nulldelimiterspace}\!\lower0.7ex\hbox{${4G}$}$,
G -- Newton's gravitation constant), the energy density of the
``equivalent'' background gravitation field (field strength $ \sim
cH_0 )$ may amount to two thirds of the critical density [11]. In
other words, such an ``acceleration field'', in principle, may be
taken as a real kinematic alternative of the dark energy.

Of course, beyond the scope of GR it is impossible to allow for
noninertiality of observer's RS consistently from relativistic
viewpoints.

\section{ Acknowledgements}

I am very grateful to Dr. V.V. Kudryashov and to Prof. A.E.
Shalyt-Margolin for their support and invaluable assistance in the
process of my work.

\newpage
\section{References}

[1] A.G. Riess et.al., Astron J. 116. (1998). 1009.

[2] S. Perlmutter et al., Astrophys. J. 517. (1999). 565.

[3] A.V. Minkevich. Phys.Lett.B, 678. (2009). 423.

[4] S.G. Turyshev. Phys.Usp. Vol.52. (2009). 1.

[5] A.G. Riess et.al., Astrophys. J. 607. (2004). 665.

[6] T. Fulton, F. Rochrich and L. Witten. Nuovo Cim., 24, (1962).
652

[7] S. Weinberg. Gravitation and Cosmology: Principle and
Applications of the General Theory of Relativity. John Wiley and
Sons, Inc., New-York-London-Sidney-Toronto. 1972

[8] M. Visser. arxiv: gr-qc/0309109 v. 4.

[9] L.M. Tomilchik. The Hubble law as a kinematical outcome of the
spase-time conformal geometry. AIP Conference Proceedings. Vol.
1205. Melville, New York 2010. 177.

[10] L.M. Tomilchik. Proc. Int. Sem. on Contemporary Problems of
Elementary Particle Physics, 17-18 January 2008. Dubna. JINR,
Russia. 2008. 194.arxiv: gr--qc /0806.0241.

[11] L.M. Tomilchik, N.G.Kembrovskaya. arxiv: gr--qc /1001.3536.

[12] G.W. Gibbons. Found.Phys. 32. (2002). 1891.

[13]
www.tapir.caltech.edu/${}_{\textrm{\symbol{126}}}$chirata/ph217/lec02.pdf

\end{document}